\documentclass{ws-ijmpc}

\begin{document}

\markboth{Casillas et al.}
{CONDENSATION IN AN ECONOMIC MODEL WITH BRAND COMPETITION}

\catchline{}{}{}{}{}

\title{CONDENSATION IN AN ECONOMIC MODEL WITH BRAND COMPETITION 
}

\author{L. CASILLAS and F.J. ESPINOSA
}

\address{Escuela de Ingenier\'{\i}a, Universidad Marista de M\'erida, 
Perif\'erico norte tablaje 13941 carretera M\'erida-Progreso\\
M\'erida, Yucat\'an,
M\'exico\\}

\author{R. HUERTA-QUINTANILLA and M. RODRIGUEZ-ACHACH}

\address{Departamento de f\'{\i}sica aplicada, Cinvestav, 
Km. 6 carretera antigua a Progreso\\
M\'erida, Yucat\'an 97310, M\'exico\\
achach@gema.mda.cinvestav.mx}

\maketitle

\begin{history}
\received{Day Month Year}
\revised{Day Month Year}
\end{history}

\begin{abstract}
We present a linear agent based model on brand competition.
Each agent belongs to one of the two brands and interacts
with its nearest neighbors. In the process the agent can decide to
change to the other brand if the move is beneficial.
The numerical simulations show that the systems always
condenses into a state when all agents belong to a single
brand. We study the condensation times for different parameters
of the model and the influence of different mechanisms to
avoid condensation, like anti monopoly rules and brand fidelity.

\end{abstract}

\ccode{PACS Nos.: 02.50.Ng, 89.65.Gh}

\section{Introduction}

During the last few years a great amount of work from the physics
community has been devoted to the application of methods from
statistical physics to the study of economic problems, or {\em
econophysics} \cite{1,2,3,4}. A wide range of problems have been
studied, for example, the shape of the distribution of price changes
\cite{5}, theoretical models of financial markets \cite{6,7}, portfolio
selection and optimization \cite{8,9}, and wealth distribution
\cite{10,11,12}, to name just a few.

Here, our interest is the study, via Monte Carlo simulations
on an agent based model, of brand competition in a bipartite
market. We present a very simple ``toy" model describing a situation
where companies can choose one of two available brands of a certain
product or service. The decision of which of the two available
options will be selected by the company is based on a 
cost and payoff mechanism. It is assumed that the product or
service has a certain cost, and the company will obtain a benefit
or payoff from its use. The products are subject to technological
innovation, and this innovation can persuade a company to
switch to the competing brand. We also explore the effect of
anti monopolistic policies and brand fidelity.

\section{The model}

Consider a population of $N$ agents that lie in a one-dimensional
chain with periodic boundary conditions. Only nearest-neighbor
interactions are allowed to keep the model as simple as
possible. Each agent represents a company that uses some kind
of tool or service of a certain brand (i.e. Apple or PC
computers, Citibank or Lloyd's financial services, etc.). 
It can be assumed that an agent will not switch to another
brand unless something is gained with the change, for example,
a company may be happy using Apple computers but, if a big improvement
or technological advance in PCs enters the market, then it may
have a reason to switch computer brands. 

In our model, we consider 
that only two different brands for a product or service exist, say
A and B. Agents are given initially one or the other brand at
random. Now suppose that a technological advance appears in one
of the two brands, for example in A. One of the agents using 
brand A will now have a better product, and the neighbors of
this agent may also want to switch to the new improved A brand, however,
this change will have an associated cost. This cost will depend on
whether the neighbor is also using brand A or is using the other
brand. Typically the cost of upgrading the same brand is less
than the cost of changing to a different brand. It may be the case
that the neighbor does not want to change because the technological
improvement is too small to compensate for the cost of upgrading
or changing, and it may decide to wait for another version
of the product to make the change. If the neighbor decides
to upgrade or change, then its own neighbor may also want
to do the same, and so on, until all agents change/upgrade or
one decides not to do so and the process stops. At this point a
new improvement in one of the brands enters the market and
the process is repeated. The simulation therefore proceeds 
as follows:

A set of $N$ agents is created. 
Each agent is assigned at random
a number that identifies the brand that is using (0 or 1). The agent
has also a real number $T$ assigned that is a measure of how new
its product is. For example an agent with a Pentium 4 computer
will have a higher value of $T$ than an agent with a Pentium 2. 
A cost $C$ of upgrading is defined, this cost applies when
an agent wants to upgrade to a new version of the same brand.
Another cost $S$ is defined that applies when an agent wants
to switch brands. Then:

\begin{enumerate}

\item Randomly pick one of the agents, say agent $i$, and increase 
its technological level $T$ by a random amount $\delta$ between 0 and 1.

\item For the neighbor $i+1$ calculate the
quantity $z=T_i - T_{i+1} - \mbox{cost}$, where cost is $C$
if agents $i$ and $i+1$ use the same brand, and $S$ if they
use different brands.

\item If $z>0$ then the gain in technological level
is greater than the cost and the agent $i$ upgrades or changes.
If $z\le 0$ then agent $i$ stays with its brand and does not upgrade
nor change.

\item Repeat steps (2) and (3) for $i+2,\, i+3, \dots$ until one of the agents
has $z<0$ in which case the avalanche of upgrading/changing stops.

\item Repeat steps (2) to (4) for neighbor $i-1$.

\item One time step has finished. Return to step (1).

\end{enumerate}

The repetition of the above process $N$ times constitutes one
Monte Carlo simulation (MCS), and its the unit of time.
Note that an agent that upgrades or switches will earn the quantity
$T_i - T_{i+1}$  but it has to pay a cost $C$ or $S$. Only when
this quantity is greater than the cost will the agent upgrade
or switch. Therefore several time steps may be necessary until
the sufficient quantity accumulates to offset the cost.

The simulation stops when all agents have the same brand. This
phenomenon of condensation is observed to happen for any
set of model parameters. The condensation time may be small or
large, depending on those parameters, but the end result
is always a single brand system. 
The brand that wins in the end is basically the one that,
by chance,
starts growing earlier, initiating a positive-feedback effect. The 
initial growth is generally triggered
by a high technological innovation. 
This condensation on one of two initially equivalent brands
has also been observed in previous works in economics \cite{13},
Ising models, and opinion dynamics \cite{14}.

\section{Results}

\begin{figure}[ph]
\centerline{\psfig{file=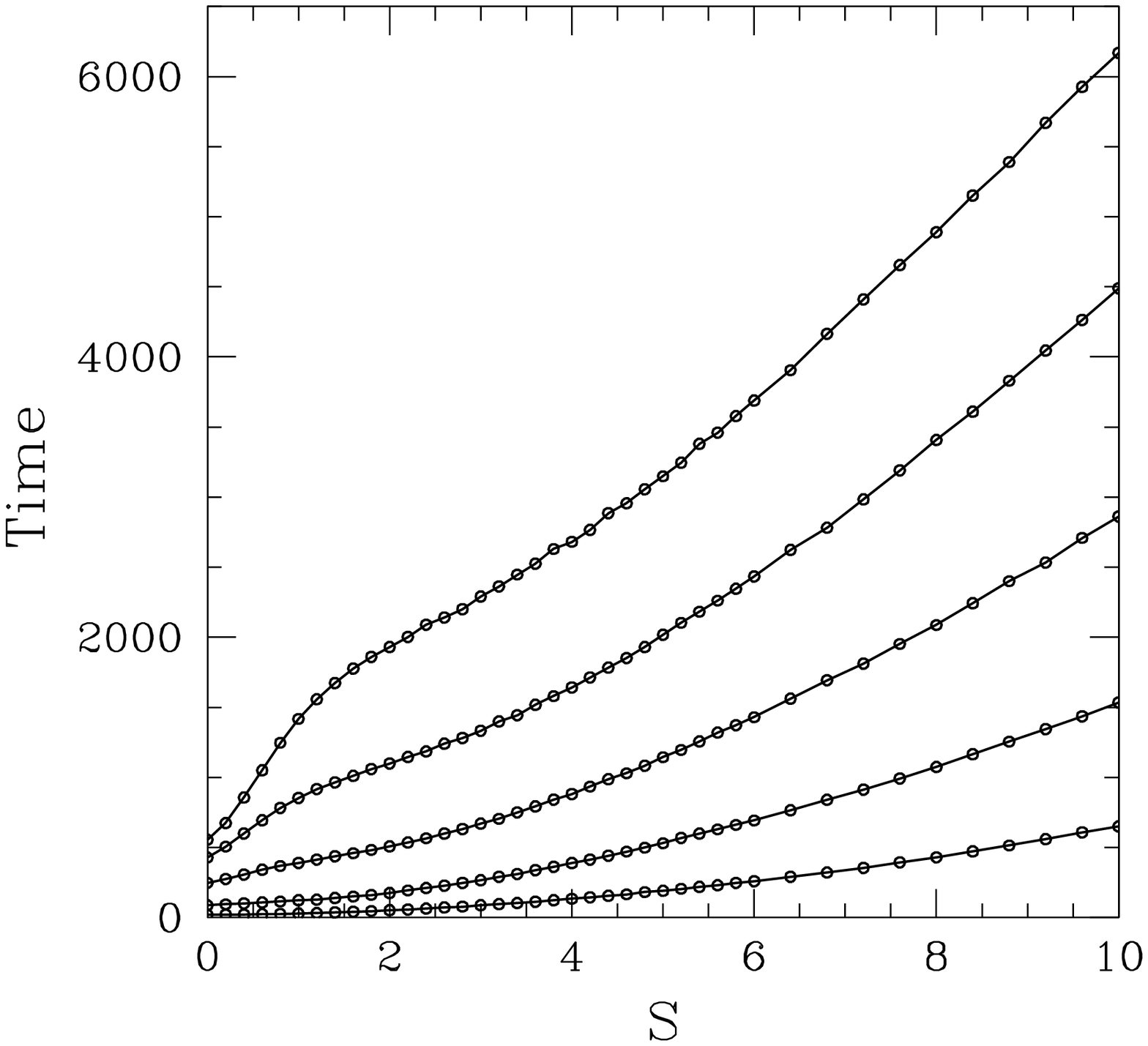,width=6.7cm}}
\vspace*{8pt}
\caption{Condensation time is plotted against the switching
cost $S$. Each curve is for a different upgrading cost $C$ which,
from bottom to top are $C=1,2,3,4,5$. Each point in the curves is
averaged over at least 2500 independent realizations.
\label{achach1}}
\end{figure}

Condensation times depend on both the cost of upgrading $C$
and the cost of switching $S$. In Fig.~\ref{achach1} we show
the results for condensation time as function of the
switching cost. Each curve is for a different value of
upgrading cost. As it can be seen, times increase for
higher values of the switching cost $S$. This is expected
since more time steps need to pass until the earnings are
high enough to offset a high cost $S$. If we increase the
upgrading cost $C$, a similar curve is obtained, only with
higher times.

\begin{figure}[ph]
\centerline{\psfig{file=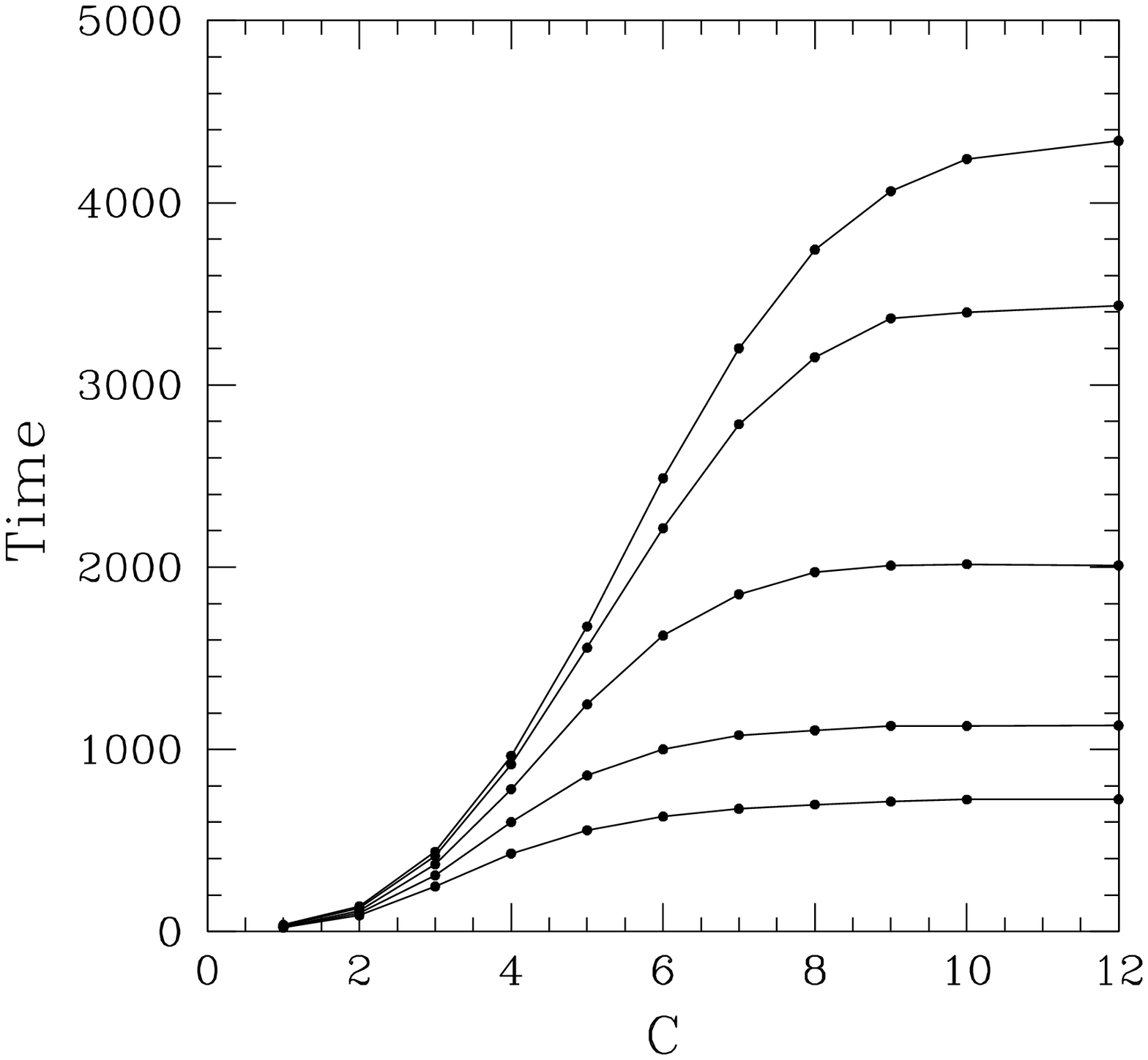,width=6.7cm}}
\vspace*{8pt}
\caption{Condensation time is plotted against the upgrading
cost $C$. Each curve is for a different switching cost $S$ which,
from bottom to top are $S=0.0,0.4,0.8,1.2,1.4$. Each point in the curves is
averaged over at least 2500 independent realizations.
\label{achach2}}
\end{figure}

The upgrading cost, unlike the switching cost, has a saturation
value. That is, if we increase $C$ while keeping $S$ fixed,
the condensation time will stop increasing at some point.
This is due to the fact that if the upgrading cost is too high compared
to the switching cost, agents will prefer to switch their brand
as soon as they can (when one of its neighbors begin using
a different brand). This is illustrated in Fig.~\ref{achach2}

Since 
there is no preference for any of the brands, either one can dominate
the market with equal probability. As we said before, the process
by which one of them grows and eventually condensates is
started by the formation of a cluster and the eventual growth and
coagulation with other same-brand clusters.

\subsection{Anti monopolistic policies}

Companies can grow too big if they are left free to develop in an
unregulated market. In order to prevent the formation of 
monopolies, several types of antitrust laws are used. In our
model, we introduce a simple antitrust rule: at each Monte Carlo
step, companies are counted and the upgrading cost of the
larger company is increased in proportion to its size. For
example, if the number of companies using product A is larger than
the number of those using B, then the upgrading cost $C$ for
companies using product A is increased by a factor

\begin{equation}
f = 1 + k\left(\frac{N_A}{N_B}-1\right),
\end{equation}
where $N_i$ is the number of companies using product $i$ and
$k$ is a parameter that controls the intensity of the antitrust
mechanism. In the above formula, $f=1$ when $N_A=N_B$, and
increases linearly as $N_A/N_B$ grows. 

\begin{figure}[ph]
\centerline{\psfig{file=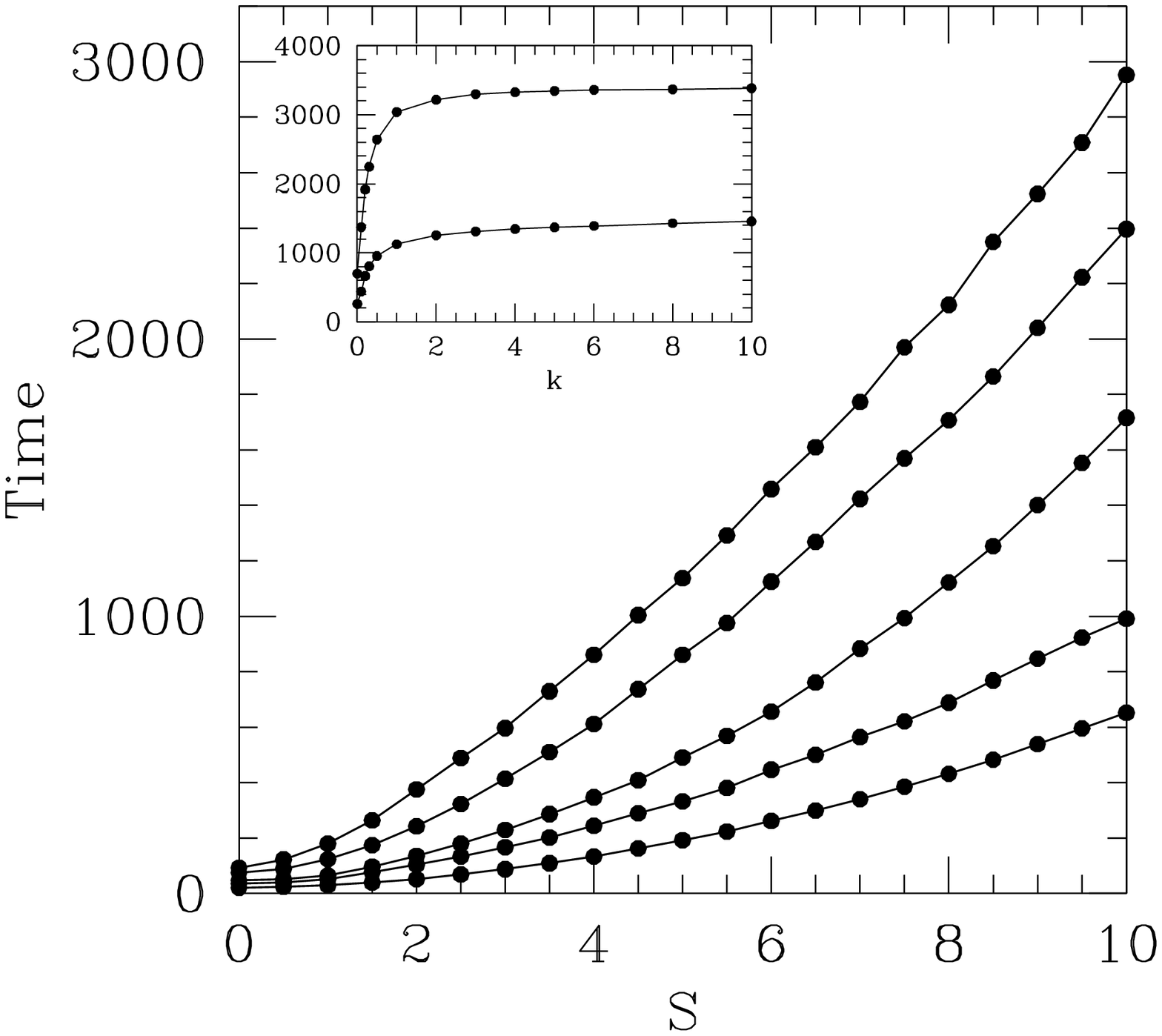,width=6.7cm}}
\vspace*{8pt}
\caption{Condensation time against the switching
cost $S$ for a fixed value of upgrading cost $C=1$. Each curve is for a 
different value of the parameter $k$ which,
from bottom to top are $k=0,0.1,0.2,1,10$. The inset shows
the condensation time as function of $k$ for a fixed value of $S=6$, with
the lower curve corresponds to a value of $C=1$ and the upper
curve is for $C=2$. Each point is
averaged over at least 2500 independent realizations.
\label{achach3}}
\end{figure}

As it can be seen from the curves in Fig.~\ref{achach3}, a higher
value of the antitrust parameter $k$ extends the condensation time, 
however, it saturates at some value of $k$, as shown in the
inset. When $k > 0$, agents are penalized for upgrading their
brand, it becomes cheaper to switch, and this retards the
condensation of the system. However, there is a point when
$k$ is so high, that agents do not upgrade, they simply stay
with their old technology for very long time. At this point
the value of $k$ no longer affects the dynamics, and we see the 
saturation shown in the inset. From a practical point of
view, this implies that an antitrust mechanism can be
good at preventing a monopoly but, if it gets too high, it
will only retard the technological innovations.

\subsection{Brand fidelity}

Another mechanism that can slow the condensation process is brand
fidelity. A company that is used to a certain product will be reluctant
to change for several reasons, for example, it will have to  train 
the employees in the use of the new product or service. 
There may be also concerns about the quality of customer
service of the new product, personal reasons, etc. All this
ends up in that the company may have a certain degree of fidelity
for a particular product or service. This mechanism is modeled
introducing a modified switching cost:

\begin{equation}
\label{fideq}
S(y)=S+\frac{y}{1+\exp(-t/10)}-\frac{y}{2},
\end{equation}
where $t$ is the number of time steps that a company
has been using a particular brand, and $y$ is a parameter
that controls the degree of fidelity. 
In this way, the cost of switching brands increases with
time, up to an asintotic value $y/2$. Since as time passes it
becomes increasingly costly for a company to change its brand, this 
mechanism implies a higher degree of fidelity for companies
that have more time using a brand. 
Note that when $t=0$, Eq.~(\ref{fideq})
reduces to $S(y)=S$.

\begin{figure}[ph]
\centerline{\psfig{file=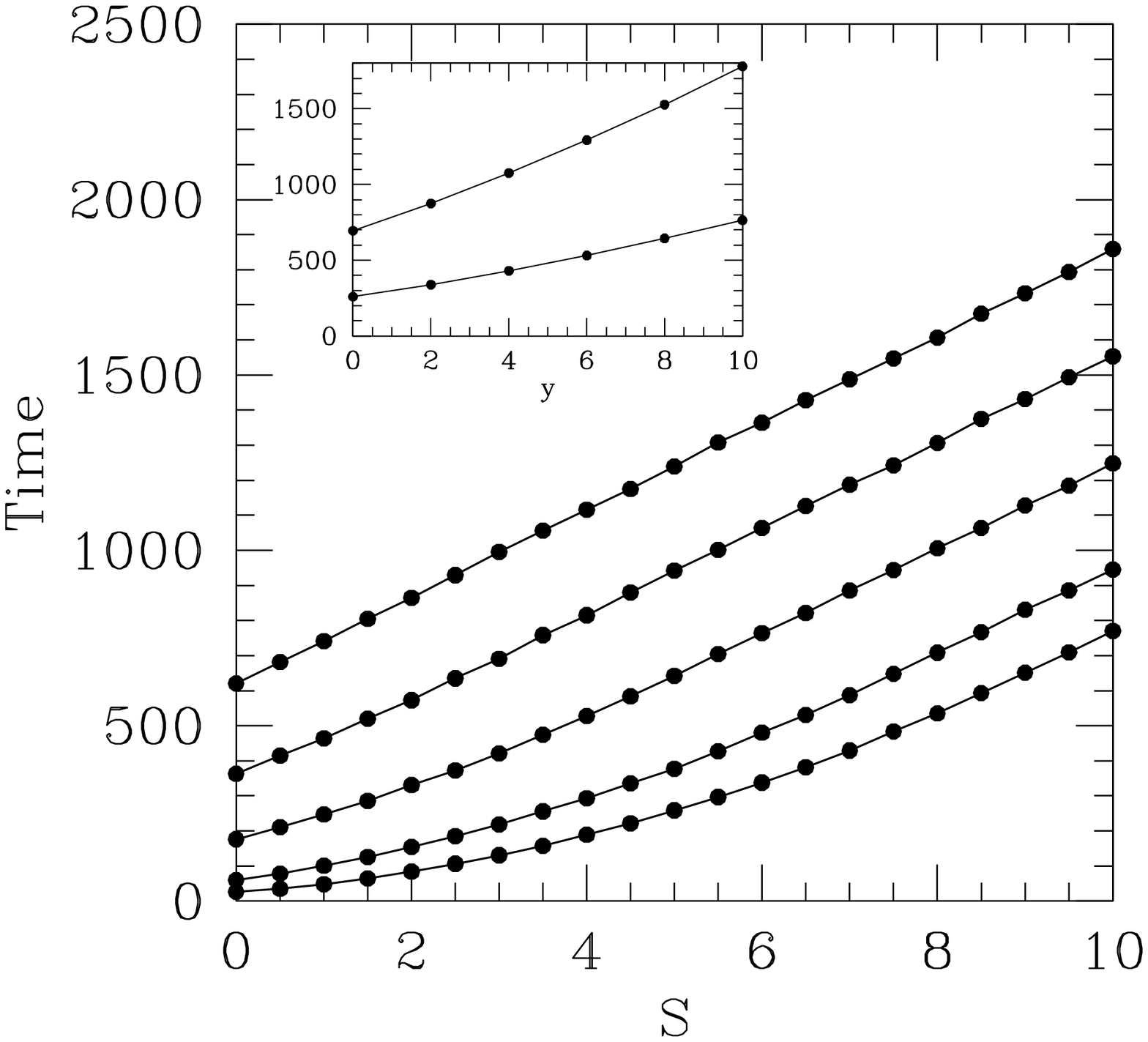,width=6.7cm}}
\vspace*{8pt}
\caption{Condensation time against the switching
cost $S$ for a fixed value of upgrading cost $C=1$. Each curve is for a 
different value of the parameter $y$ which,
from bottom to top are $y=2,5,10,15,20$. The inset shows
the condensation time as function of $y$ for a fixed value of $S=6$, with
the lower curve corresponds to a value of $C=1$ and the upper
curve is for $C=2$. Each point is
averaged over at least 2500 independent realizations.
\label{achach4}}
\end{figure}

Figure \ref{achach4} shows the results of this mechanism. Again we see
that condensation time increases with the fidelity parameter,
in almost a linear fashion. The inset of the figure shows that,
unlike the antitrust parameter of the previous section, the
fidelity parameter can be augmented without saturation in the
condensation time. Of course, the extreme case when a company
is completely faithful to a brand will make the
condensation time to diverge, therefore no saturation is
expected in this parameter.

\section{Conclusions}

We have studied a model of brand competition in the simplest case
of two brands and nearest-neighbor one-dimensional dynamics. 
It is found that the system always reaches a condensed state
where a single brand dominates the market. This state is reached
despite the introduction of mechanisms to avoid it, like anti monopoly
rules and fidelity to the brand behavior. These mechanisms succeed in
delaying the condensation but do not prevent it. The final 
single-brand state cannot be avoided and the system is doomed
to become a monopoly. We have cited results
from other authors \cite{13,14} that show a similar behavior, even in more than
one dimension. In the Sznajd model
of opinion dynamics for example, condensation appears in all dimensions.
Our future work will therefore extend the present model to include multi-brand
and two-dimensional situations.

\section*{Acknowledgments}
This work was partially supported by a grant from CONACyT (M\'exico).

\end{document}